\documentclass{vldb}
\usepackage{graphicx}
\usepackage{balance}  
\usepackage{enumitem}
\usepackage{xcolor}
\usepackage[nolist,nohyperlinks]{acronym}
\usepackage{hyphenat}
\usepackage{algorithm}
\usepackage[noend]{algpseudocode}
\usepackage{booktabs}
\usepackage{pgfplots}
\usepgfplotslibrary{groupplots}
\usetikzlibrary{matrix}
\usepackage{tikz}
\usepackage{booktabs} 				
\usepackage{amssymb}

\PassOptionsToPackage{hyphens}{url}	
\usepackage{hyperref}
\usepackage[capitalise]{cleveref}	

\newlist{requirements}{enumerate}{1}
\setlist[requirements]{label*=(\arabic*), ref=\arabic*}

\crefname{requirementsi}{requirement}{requirements}
\Crefname{requirementsi}{Requirement}{Requirements}

\renewcommand{\thetable}{\arabic{table}}

\begin{acronym}
	\acro{PoW}{Proof of Work}
	\acro{SPV}{Simplified Payment Verification}
	\acro{BAR}{Byzantine, Altruistic, Rational}
	\acro{ETH}{Ether}
	\acro{EVM}{Ethereum Virtual Machine}
	\acro{DAG}{Directed Acyclic Graph}
	\acro{DApp}{decentralized application}
\end{acronym}

\vldbTitle{Testimonium: A Cost-Efficient Blockchain Relay}
\vldbAuthors{Philipp Frauenthaler, Marten Sigwart, Christof Spanring, and Stefan Schulte}
\vldbDOI{https://doi.org/10.14778/xxxxxxx.xxxxxxx}
\vldbVolume{12}
\vldbNumber{xxx}
\vldbYear{2019}
\newcommand\Mark[1]{{\large\textsuperscript#1}}

\begin{document}


\title{Testimonium:\\A Cost-Efficient Blockchain Relay}



%
%
%
%

\numberofauthors{4} 

\author{
	Philipp Frauenthaler\Mark{1}, Marten Sigwart\Mark{1}, Christof Spanring\Mark{2}, Stefan Schulte\Mark{1}\\[2pt]
	\affaddr{\Mark{1}TU Wien, Vienna, Austria, \Mark{2}Pantos GmbH, Vienna, Austria}\\[2pt]
	\affaddr{\Mark{1}\{p.frauenthaler, m.sigwart, s.schulte\}@dsg.tuwien.ac.at, \Mark{2}contact@pantos.io}
}

\maketitle

\begin{abstract}
Current blockchain technologies provide very limited means of interoperability. In particular, solutions enabling blockchains to verify the existence of data on other blockchains are either very costly or are not fully decentralized.

To overcome these limitations, we introduce Testimonium, a novel blockchain relay scheme that applies a validation-on-demand pattern and the on-chain execution of Simplified Payment Verifications to enable the verification of data across blockchains while remaining fully decentralized. Evaluating the scheme for Ethereum-based blockchains shows that Testimonium achieves a cost reduction of up to 92\% over existing solutions. As such, the scheme lays a strong foundation for generic blockchain interoperability. For instance, it enables the development of an atomic-commit protocol for distributed transactions across blockchains.

\end{abstract}

\section{Introduction}
\label{sec:intro}
For its ability to store data in a decentralized and immutable way, blockchain technology has gained much attention by the industry and research communities as potentially disruptive in areas such as finance~\cite{nakamoto2008bitcoin}, business process management~\cite{prybila2017runtime}, data provenance~\cite{ruan2019fine, sigwart2019blockchain}, supply chain management~\cite{tian2016agri}, or healthcare~\cite{mettler2016blockchain}. To take the diverse requirements of these use cases into account, a variety of different blockchain platforms have been developed~\cite{yli16}, not unlike the emergence of various NoSQL databases as alternatives to traditional database systems~\cite{srinivasan2011nosql}. In this field of multiple independent and unconnected blockchains~\cite{schulte2019towards}, it is unlikely that a ``blockchain to rule them all'' emerges~\cite{zamyatin2019sok}. This reinforces the need for interoperability solutions, especially in scenarios where organizations which utilize different blockchains collaborate with each other.

In such scenarios, it may be essential that state changes across blockchains are treated as an atomic unit which either succeeds or fails~\cite{arun2019cross-chain-comm}. However---despite already being well-established for traditional databases~\cite{lampson1993new}---such distributed transactions cannot be seamlessly transferred to the blockchain field, since any interaction with external systems might jeopardize the integrity and decentralization guarantees of blockchains like Bitcoin and Ethereum. Ideally, blockchain interoperability is achieved while preserving the properties of integrity and decentralization, i.e., the underlying cross-blockchain communication should not rely on trust in a centralized party.

One blockchain interoperability approach that is particularly promising are so-called relay schemes. Relay schemes replicate block information of some source blockchain within a destination blockchain to allow the latter to verify the existence of data~(e.g., transactions) on the source blockchain without requiring trust in a centralized entity~\cite{buterin2016interoperability}. The ability to verify arbitrary data across blockchains paves the way for more generic blockchain interoperability, e.g., by enabling the implementation of an atomic-commit protocol for distributed transactions across multiple blockchains~\cite{herlihy19cross}. 

For these verifications to be trustworthy, the block information of the source blockchain needs to be validated by the destination blockchain according to the validation rules of the source blockchain. However, depending on the source and destination blockchains, this validation can be very expensive when being performed on-chain. Within current relay solutions~\cite{btcrelay, luu2017peacerelay}, this inevitably leads to either very high operational cost or, if the expensive on-chain validation is by-passed, the need to rely on a centralized component.

To overcome this issue, we introduce Testimonium, a relay scheme that is fully decentralized while being cost-efficient even for blockchains with expensive validation protocols. The key to this concept is a sophisticated incentive scheme combined with a validation-on-demand approach. We evaluate the scheme in a proof of concept implementation for Ethereum-based blockchains to show that it achieves a cost reduction of up to 92\% over existing relay solutions.

The paper is organized as follows. \Cref{sec:background} provides background information while \cref{sec:design} describes the technical contributions of this paper. In \cref{sec:evaluation}, we evaluate the proposed relay with regards to security and operational cost. In \cref{sec:related}, we give an overview of related work. Finally, \cref{sec:conclusion} concludes the paper.\\\\

\section{Background}
\label{sec:background}
This section discusses important background information for Testimonium. For this, we first explain the concept of \ac{SPV}. We then describe how \acp{SPV} are used to facilitate blockchain relay schemes.

\subsection{Simplified Payment Verification}
\label{sec:spv}


\begin{figure}
	\centering
	\includegraphics[width=0.85\linewidth]{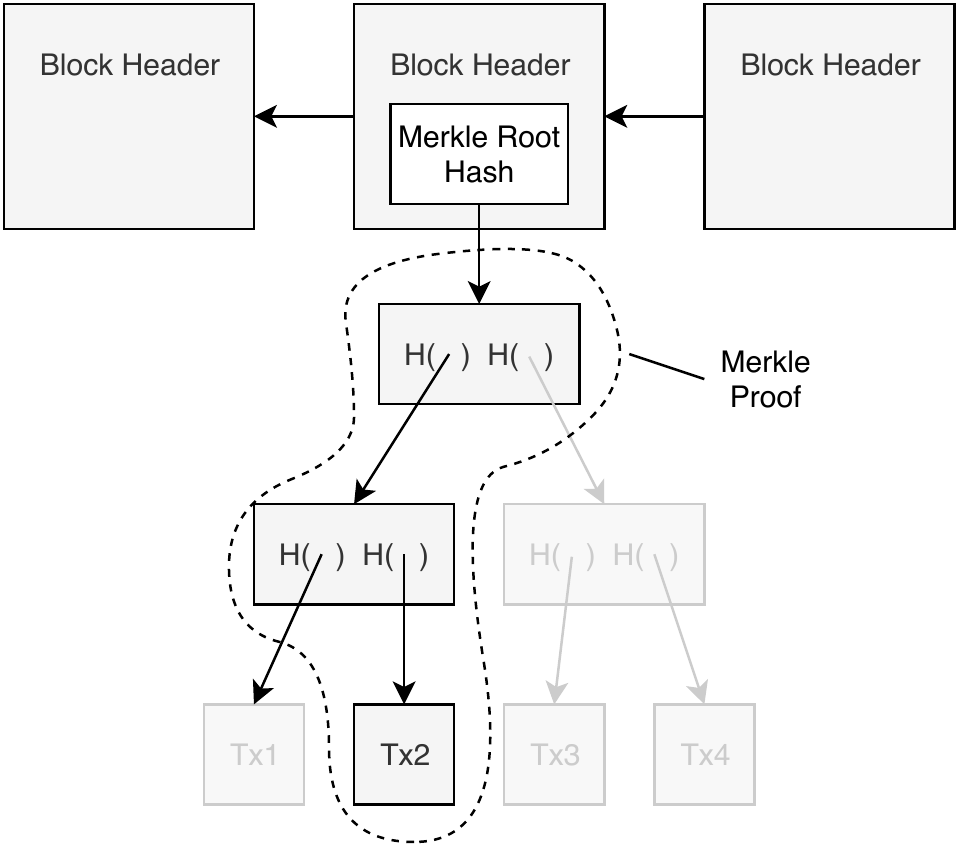}
	\caption{Block header with Merkle tree and corresponding Merkle proof of membership for Tx2}
	\label{fig:merkle}
\end{figure}


\acp{SPV} enable clients to cryptographically verify that a particular transaction is part of a blockchain without having to store the full blockchain~\cite{nakamoto2008bitcoin}. Instead of the full blockchain, an \ac{SPV} client only needs to keep a copy of the block headers. In contrast to complete blocks, block headers only store meta data (e.g., the block number) but no transaction data. Thus, block headers only consume a fraction of the space a complete block needs. 

To get new block headers, the client can query nodes having access to the full blockchain. Once the client has a copy of the headers of the blockchain, users can prove to the \ac{SPV} client the inclusion of transactions in the blockchain and the \ac{SPV} client can verify these proofs without keeping a copy of the actual transaction data. For that, the client leverages the fact that the transactions of a block are stored as leaves in a so-called Merkle tree~\cite{narayanan2016bitcoin} and that the hash of the Merkle tree's root node (Merkle root hash) is stored in the block's header~(see~\cref{fig:merkle}).

In case a user wants to prove the inclusion of a particular transaction to an \ac{SPV} client, the user needs to provide a so-called Merkle proof of membership. This proof contains all nodes of the path from the transaction (leaf) up to the root node~(see~\cref{fig:merkle}). When receiving such a proof, an \ac{SPV} client recalculates the hashes of all nodes along the path from the leaf (i.e., the transaction) up to the root node. If the final hash matches the Merkle root hash of the stored block header, the membership of the transaction within the corresponding block has been successfully verified.

\subsection{Relay Schemes}
\label{sec:relay-schemes}
\ac{SPV} clients have the ability to verify whether or not a particular transaction exists on some blockchain. Blockchain relays like BTC Relay~\cite{btcrelay} and PeaceRelay~\cite{luu2017peacerelay} utilize this capability to enable transaction inclusion verifications across blockchains. Essentially, relays are \ac{SPV} clients for a source blockchain running on a destination blockchain. For instance, BTC Relay is a relay running on the Ethereum blockchain (i.e., the destination blockchain) enabling transaction inclusion verifications on block headers from the Bitcoin blockchain (i.e., the source blockchain).

For successful \ac{SPV}, the relay needs to know about the block headers of the source blockchain. For that, block headers of the source blockchain need to be constantly submitted to the relay by off-chain clients~\cite{buterin2016interoperability}. With knowledge of all block headers of the source blockchain, the relay can leverage \ac{SPV} to verify on the destination blockchain that a particular transaction has been included in the source blockchain.

To keep the system fully decentralized, i.e., to not require any trust in any off-chain client, newly submitted block headers are first validated by the relay before transaction inclusion verifications can be performed on them~\cite{buterin2016interoperability}. Furthermore, competing branches of a blockchain are a common occurrence especially in \ac{PoW} blockchains~\cite{wood2014ethereum}. While these branches usually consist of valid blocks, only one branch is eventually accepted as the main chain (e.g., in \ac{PoW} blockchains, the branch with the greatest amount of work invested in it~\cite{buterin2016interoperability}). As transaction inclusion verifications should only be performed for block headers that are part of the current main chain of the source blockchain, the relay needs to track the branch representing the main chain.

A block header is considered valid if it complies with the source blockchain's standard validation procedure. Among other things, this usually involves validating the consensus algorithm. For instance, in \ac{PoW} it needs to be verified that enough work has been performed whereas for other blockchains it may consist of checking that at least a certain amount of validators (e.g., 2/3) signed the block~\cite{buterin2016interoperability}.

Existing relays like BTC Relay perform the source blockchain's header validation for every submitted block header. However, depending on the source and destination blockchains, verifying the validity of a block header on-chain can be computation- and storage-intensive. For example, validating Ethash, the \ac{PoW} algorithm of Ethereum, requires fragments of the data used during mining to be available on-chain. Even optimized solutions need approximately 3~million gas when executed on an Ethereum-based blockchain (see \cref{sec:evaluation}). Performing this validation for every block header of the source blockchain leads to extremely high operational cost. To the best of our knowledge, current relay schemes offer no solutions to this problem without involving trusted third parties (see \cref{sec:related}).

To tackle this issue, we introduce Testimonium, a relay scheme that achieves a significant cost reduction over traditional blockchain relays. The fundamental concepts of Testimonium are discussed in the following section.

\section{Testimonium Relay Scheme}
\label{sec:design}
This section introduces Testimonium, a relay scheme that keeps cost of executing \acp{SPV} on-chain to a minimum by deploying a validation-on-demand pattern for relayed block headers. Testimonium requires no trust in a single entity as validations are executed on-chain with a reward structure incentivizing participation.

As mentioned in \cref{sec:background}, blockchain relays store a copy of the block headers of some source blockchain. The Testimonium relay scheme assumes the block headers of the source blockchain to be based on the data structure proposed by Satoshi Nakamoto~\cite{nakamoto2008bitcoin}, which is for example the case for Bitcoin and Ethereum, and the many forks of these blockchain protocols. That is, block headers contain at least the hash of the block's parent, the block's height, and the hash of the root node of the Merkle tree containing the block's transactions. These three fields are referred to as \emph{parentHash}, \emph{blockHeight}, and \emph{merkleRoot}, respectively.
 In \cref{sec:bc-requirements}, we discuss additional requirements the involved blockchains need to satisfy. Furthermore, Testimonium introduces some additional fields typically not present in block headers. These fields are needed for executing certain actions and are prefixed with an~\emph{m} (for ``meta'') to distinguish them from fields usually present in a block header.

Testimonium consists of the relay itself (an on-chain program running on the destination blockchain) and two types of off-chain clients: submitters are responsible for relaying block headers from the source blockchain to the destination blockchain, and disputers are responsible for detecting and disputing submitted illegal block headers.

\subsection{Replicating the Source Blockchain}
\label{sec:replicate-source-chain}
 In relay schemes, off-chain clients continuously submit block headers of the source blockchain to the relay on the destination blockchain. When a new header is submitted, multiple actions are performed by the Testimonium relay before the submitted header can be used for verifying the existence of transactions on the source blockchain.
 
 \cref{alg:submit} shows the pseudo code for the procedure undertaken by the Testimonium relay whenever a new block header is submitted by an off-chain client. Right after the arrival of a new header, it is checked whether the retrieved header has already been submitted to the relay (Line~\ref{alg:submit:check-hash}). If this is the case, the submitted header is rejected. Submitted block headers are stored in a global hashmap using the header's hash as key and the header itself as value. The function \Call{hash}{} represents the hash function used on the source blockchain to calculate block hashes, e.g., when storing Bitcoin headers, \Call{hash}{} would implement \emph{SHA-256}~\cite{narayanan2016bitcoin}.
 
 Next, it is checked whether the block referenced by the field \emph{parentHash} exists on the relay (i.e., whether it has already been submitted to the Testimonium relay), as shown in Line~\ref{alg:submit:check-parent}. This ensures that only a continuous chain of block headers is replicated within the Testimonium relay.
 
 


\begin{algorithm}[t]
	\caption{Procedure performed by the Testimonium relay when receiving a new header of the source blockchain}
	\label{alg:submit}
	\begin{algorithmic}[1]
		\Function{SubmitBlockHeader}{\emph{header, submitter}}
		\If {\emph{headers}.contains(\Call{hash}{\emph{header}}) == \emph{true}}\label{alg:submit:check-hash}
		\State \Return \emph{false}
		\EndIf
		\State \emph{parentHash} = \emph{header.parentHash}
		\If {\emph{headers}.contains(parentHash) == \emph{false}}\label{alg:submit:check-parent}
		\State \Return \emph{false}
		\EndIf
		\State \emph{headers}.put(\Call{hash}{\emph{header}}, \emph{header}) \label{alg:submit:add-header}
		\State \emph{header.m.lockedUntil} = \emph{now} + \emph{LOCK\_PERIOD} \label{alg:submit:lockperiod}
		\State \emph{header.m.submitter} = \emph{submitter} \label{alg:submit:submitter}
		\State \emph{parent} = \emph{headers}.get(\emph{parentHash}) \label{alg:submit:parent}
		\State \emph{parent.m.chldn}.append(\Call{hash}{\emph{header}}) \label{alg:submit:succs}
		\State \emph{branchHeads}.add(\Call{hash}{\emph{header}}) \label{alg:submit:addep}
		\If {\emph{branchHeads}.contains(\emph{parentHash})} \label{alg:submit:if-parent-ep}
		\State \emph{branchHeads}.remove(\emph{parentHash}) \label{alg:submit:remove-parent}
		\State \emph{header.m.branchId} = \emph{parent.m.branchId} \label{alg:submit:forkIdx-pforkIdx}
		\State \emph{header.m.junction} = \emph{parent.m.junction} \label{alg:submit:lfork-plfork}
		\Else
		\State \emph{lastBranchId} = \emph{lastBranchId} + 1 \label{alg:submit:lastForkIdx}
		\State \emph{header.m.branchId} = \emph{lastBranchId} \label{alg:submit:forkIdx-lastForkIdx}
		\State \emph{header.m.junction} = \emph{parentHash} \label{alg:submit:lfork-parent}
		
		\If {\emph{parent.m.chldn.length} == 2} \label{alg:submit:branch-first-time}
		\State \Call{setJunction}{\emph{parent.m.chldn}[0],\emph{parentHash}}\label{alg:submit:set-latest-fork-branch}
		\EndIf
		\EndIf
		\State \emph{mainChainHead} = \Call{getMainChainHead}{\,} \label{alg:submit:valid-fork}
	
		\EndFunction
	\end{algorithmic}
\end{algorithm}

\subsubsection*{Optimistically Accepting Block Headers}
As stated in \cref{sec:relay-schemes}, executing the source blockchain's header validation procedure on the destination blockchain can be very expensive. 
Performing this validation for every block header of the source blockchain would lead to high operational cost. Therefore, Testimonium follows an optimistic approach where received block headers are accepted at first without being fully validated (Line~\ref{alg:submit:add-header}). 

Of course, since submitted block headers are not fully validated, illegal block headers may be accepted by the Testimonium relay, potentially enabling the verification of illegal transactions. Testimonium prevents this by assigning a lock period to newly received block headers~(Line~\ref{alg:submit:lockperiod}). During this period, a block header is considered ``locked'', meaning that it cannot be used for transaction inclusion verifications. Furthermore, within this lock period, off-chain clients (i.e., the \textit{disputers}) can dispute headers they deem illegal. In case of a dispute, the header validation is carried out by the Testimonium relay and if the validation fails, the illegal block header is eliminated. \cref{sec:disputing-headers} discusses the disputing of illegal block headers in detail. Further, to make submitters of illegal block headers accountable, the submitter's address is stored in the received block header by using the field \emph{m.submitter}, as shown in Line~\ref{alg:submit:submitter}.

\subsubsection*{Handling Blockchain Branches}
\label{sec:handling-forks}
In \ac{PoW} blockchains like Bitcoin and Ethereum, multiple valid blocks with the same block height can exist in parallel, forming so-called blockchain branches. While multiple branches can exist in parallel, only one of these branches represents the current main chain of the blockchain, e.g., in \ac{PoW} blockchains this is the branch with the most amount of work invested in it~\cite{buterin2016interoperability}. As more block headers are appended to branches, the main chain of a blockchain may change over time. This represents a challenge to on-chain \ac{SPV} execution since transaction inclusion verifications should only be successful if the requested block is part of the main chain.

A branch head represents the most recent block header of a blockchain branch. To keep track of all existing branches of the source blockchain, the Testimonium relay tracks the head of each branch. This way, the relay is able to constantly re-evaluate the current main chain of the source blockchain, e.g., in \ac{PoW} blockchains, the main chain is identified by searching for the branch with the highest total difficulty. 

Whenever a new block header has passed the validation, it is added to the set of branch heads~(Line~\ref{alg:submit:addep}) as it either becomes the new head of an existing branch or it represents the start of a completely new branch. If a block header continues an already existing branch, it replaces its parent as a branch head~(Lines~\ref{alg:submit:if-parent-ep}ff). If a block header creates a new branch, the header is merely appended to the branch head set as the creation of a new branch does not affect existing branch heads. The branch head of the main chain is stored in the global variable \emph{mainChainHead}. Whenever a new block header is submitted, the current main chain is re-evaluated~(Line~\ref{alg:submit:valid-fork}).


\subsubsection*{Enabling Efficient Verifications}
Whenever a transaction inclusion verification is requested on a certain block, the Testimonium relay needs to determine whether the block is part of the main chain of the source blockchain. As each block header contains a hash pointer to its parent forming a linked list, we could simply trace from the main chain's head (stored in variable~\emph{mainChainHead}, Line \ref{alg:submit:valid-fork}) all the way back until we reach the requested block or the genesis block. However, depending on how far back the requested block lies, this traversal can be expensive. To make this traversal more cost-efficient, the Testimonium relay stores additional helper variables with each block header.

First, it stores a reference to the preceding branch junction, i.e., a reference to the header where the branch of the newly submitted block header branched off. Second, it stores a number identifying a subpath of the current branch, i.e., the branch of which the newly submitted block header is the new head. We refer to this number as branch id. This meta data is stored in the fields \emph{m.junction} and \emph{m.branchId}, respectively. In case the submitted header continues an existing branch, \emph{m.junction} and \emph{m.branchId} are set to the corresponding field values of its parent, as shown in Lines~\ref{alg:submit:forkIdx-pforkIdx} and \ref{alg:submit:lfork-plfork}. If a new branch occurs, \emph{m.branchId} gets assigned the maximum branch id used so far incremented by one and \emph{m.junction} of the submitted header is set to the parent's hash~(Lines~\ref{alg:submit:lastForkIdx} to~\ref{alg:submit:lfork-parent}). The helper fields \emph{m.branchId} and \emph{m.junction} enable a more efficient search when verifying transaction inclusions, as the backwards traversal can be executed in jumps from branch junction to branch junction rather than from block to block.

Furthermore, the Testimonium relay keeps track of each block header's children. We refer to some block~$c$ as child of some other block~$p$ if the \emph{parentHash} of $c$ holds a reference to $p$. Further, we refer to a block $d$ as descendant of some other block $p$ if $d$ can reach $p$ using the hash pointers (\emph{parentHash}) forming the linked list. Analogous, we refer to block $p$ as predecessor of block $d$. Whenever a new block header is received, the Testimonium relay adds its hash to its parent's child list~(Line~\ref{alg:submit:succs}). Typically, a header only has one child. If the header is a branch junction, the list contains at least two children (i.e., the hashes of the block headers branching off from the header). 

The child list is useful for a number of reasons. First, it allows an easier updating of the \emph{m.junction} field. In case a new branch junction emerges~(Line~\ref{alg:submit:branch-first-time}), the \emph{m.junction} field of the descendants of the new branch junction needs to be set to the hash of the branch junction. The function \Call{setJunction}{} (Line~\ref{alg:submit:set-latest-fork-branch}) updates each descendant until a header is reached that has either no child~(i.e., the header is a branch head) or at least two children (i.e., the header represents another branch junction).  
Second, in case a block header is successfully disputed, the child list is used to delete all block headers that were appended to the illegal block (see \cref{sec:disputing-headers}). 

Finally, the child list is also used to facilitate transaction inclusion verifications since it helps to determine the number of confirming blocks. In the next section, we look at how these verifications are carried out in detail.

\subsection{Transaction Inclusion Verifications}
\label{sec:verifying-tx}


\begin{figure*}[t]
	\includegraphics[width=\linewidth]{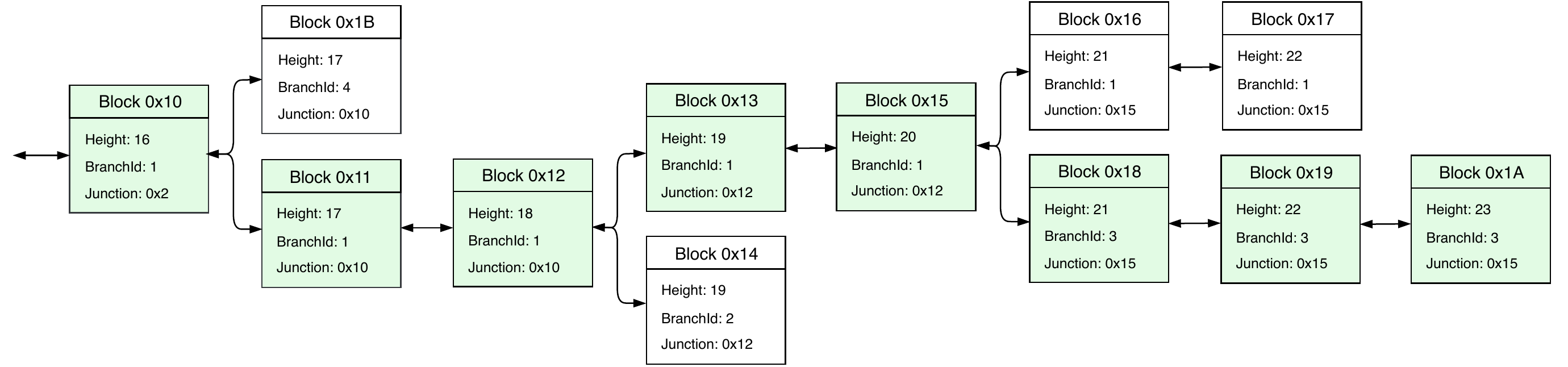}
	\caption{An example illustrating the replication of a source blockchain within a Testimonium relay. Headers are double-linked, denoted by arrows pointing in both directions. Green headers represent the current main chain of the source blockchain. For the sake of simplicity, block hashes are in ascending order to make it clearly evident which block headers have been submitted before others to the relay, e.g., block header 0x11 was submitted after block header 0x10, 0x1B after 0x1A, and so on. Block headers \emph{0x1B}, \emph{0x14}, \emph{0x17}, and \emph{0x1A} are heads of the corresponding branch. Block headers \emph{0x10}, \emph{0x12}, and \emph{0x15} represent branch junctions.}
	\label{fig:forks}
\end{figure*}

As soon as a block header of the source blockchain is replicated within the Testimonium relay on the destination blockchain, transaction inclusion verifications on that block header are possible. For that, a client (e.g., a smart contract) sends a request to the Testimonium relay in the form of ``Is transaction $tx$ of block $b$ part of the source blockchain and confirmed by at least $n$ blocks?''. To answer such a verification request, the relay executes an on-chain \ac{SPV}. First, it verifies that the header of block $b$ is known, unlocked (i.e., the lock period has passed), and part of the main chain of the source blockchain. Second, it verifies that block $b$ is confirmed by at least $n$ succeeding blocks. Finally, the Merkle proof of membership which has to be submitted together with the verification request is validated.

\subsubsection*{Verifying Block Membership on the Main Chain}
As outlined in \cref{sec:handling-forks}, the Testimonium relay assigns a branch id to every submitted block header. Whenever a submitted block header branches off a new branch, the currently maximum branch id is incremented by one and assigned to the new header. In case a submitted header continues an existing branch, it gets assigned the branch id of its parent. 

Consider the example illustrated in \cref{fig:forks} which shows a source blockchain replicated within a Testimonium relay. As can be seen, all block headers between two consecutive branch junctions or between a branch junction and a branch head have the same branch id \emph{BranchId}. Since a submitted header's branch id is either set to its parent's branch id or to the maximum branch id incremented by one, we know that all descendants of some block header $h$ have a branch id equal to or greater than the branch id of $h$. Analogous, all predecessors of $h$ have a branch id equal to or lower than the branch id of $h$.

Hence, when verifying the main chain membership of some block header $h$ that has a branch id greater than the branch id of the main chain's head, we know that header $h$ is not part of the main chain without having to traverse any headers of the main chain.

This constraint is used by \cref{alg:part-of-fork} to verify that a certain block header is part of the main chain of the source blockchain, i.e., whether the block header is a predecessor of the main chain's head. The parameter \emph{blockHash} holds the hash of the block header to check, whereas \emph{target} contains the corresponding header (Line~\ref{alg:part-of-fork:get-header}).

The algorithm starts at the main chain's head~(Line~\ref{alg:part-of-fork:init-current}). From that point on, the algorithm traverses each branch junction reachable from the head as long as the currently traversed block header's branch id is greater than the branch id of \emph{target} (Lines~\ref{alg:part-of-fork:loop-start}-\ref{alg:part-of-fork:loop-end}).

Since all headers between two consecutive branch junctions or between a branch junction and the branch's head always have the same branch id, it is completely sufficient to only traverse branch junctions. Since all headers in-between are skipped, the algorithm needs to check only few block headers instead of traversing all headers between the header to look for (i.e., \emph{target}) and the main chain's head.

After the execution of the while-loop, variable \emph{current} contains the branch junction with a branch id equal to or lower than the branch id of \emph{target}. If the branch id of \emph{current} is strictly lower than the branch id of \emph{target}, \emph{target} cannot be a part of the main chain and the function returns false~(Line~\ref{alg:part-of-fork:check-fork-idx}f). Otherwise, we know that both headers have the same branch id, i.e., they are both part of at least one common branch. 

However, it can still be the case that \emph{target} is not part of the main chain. Consider block headers~\emph{0x15} and \emph{0x16} in \cref{fig:forks}. If \emph{0x15} is the header returned by the loop (i.e., \emph{current}) and \emph{0x16} the header to look for (i.e., \emph{target}), \emph{0x15} is part of the main fork while \emph{0x16} is not. As such, the Testimonium relay compares the block heights of \emph{current} and \emph{target}. If \emph{target.blockHeight} $>$ \emph{current.block\-Height}, \emph{target} is not part of the main chain and the function returns false. If \emph{target.blockHeight} $<=$ \emph{current.block\-Height}, \emph{current} and \emph{target} are either the same or \emph{current} is a branch junction between \emph{target} and the main chain's head. Hence, in both cases, \emph{target} is part of the main chain.


Further, when traversing branch junctions, each branch junction is checked whether its lock period has passed. If an unlocked branch junction is encountered for the first time, its child along the path of the main chain is stored in the variable \emph{confirmStart}~(Lines~\ref{alg:part-of-fork:confirmstart-start}-\ref{alg:part-of-fork:confirmstart-end}). If the main chain's head is already unlocked, then all branch junctions on the main chain are unlocked as well (they have been submitted either before or at the same time as the main chain's head). In this case, we store the main chain's head in the variable \emph{confirmStart} before entering the loop~(Lines~\ref{alg:part-of-fork:head-is-unlocked}-\ref{alg:part-of-fork:confirm-current}), causing the check of each encountered branch junction's lock period to be skipped. Besides a boolean indicating whether the header to look for lies on the main chain, the algorithm also returns the \emph{confirmStart} variable. The header referenced by \emph{confirmStart} is used as starting point for verifying whether the requested header has enough confirming block headers. Once more, consider the example in \cref{fig:forks}. Assuming the function is called with block hash \emph{0x11}, and block \emph{0x15} is the latest unlocked branch junction, then the starting point for the confirmation verification is block \emph{0x18}. The algorithm verifying the number of confirmations is discussed in the following section. 

\begin{algorithm}[t]
	\caption{Verifies whether the block header referenced by \emph{blockHash} is part of the main chain of the source blockchain.}
	\label{alg:part-of-fork}
	\begin{algorithmic}[1]
		\Function{isPartOfMainChain}{\emph{blockHash}}
		\State \emph{target} = \emph{headers}.get(\emph{blockHash})\label{alg:part-of-fork:get-header}
		\State \emph{current} = \emph{headers}.get(\emph{mainChainHead}) \label{alg:part-of-fork:init-current}
		\State \emph{confirmStart} = 0\label{alg:part-of-fork:confirm-init}
		\If {\emph{current.m.lockedUntil} $<$ now}\label{alg:part-of-fork:head-is-unlocked}
		\State \emph{confirmStart} = \emph{mainChainHead}\label{alg:part-of-fork:confirm-current}
		\EndIf
		\While {\emph{current.m.branchId} $>$ \emph{target.m.branchId}}\label{alg:part-of-fork:loop-start}
			\State \emph{oldBranchId} = \emph{current.m.branchId}
			\State \emph{current} =  \emph{headers}.get(\emph{current.m.junction})\label{alg:part-of-fork:jmp-fork}
			\If{\emph{confirmStart} != 0}\label{alg:part-of-fork:confirmstart-start}
				\textbf{continue}
			\EndIf			
			\If{\emph{current.m.lockedUntil} $<$ now}
				\State \emph{confirmStart} = 
				\State \Call{getChildByBranch}{\emph{current}, \emph{oldBranchId}}
			\EndIf \label{alg:part-of-fork:confirmstart-end}
		\EndWhile \label{alg:part-of-fork:loop-end}
		\If {\emph{current.m.branchId} $<$ \emph{target.m.branchId}}\label{alg:part-of-fork:check-fork-idx}
		\State \Return (\emph{false}, \emph{confirmStart})
		\EndIf
		\If {\emph{current.blockHeight} $<$ \emph{target.blockHeight}}\label{alg:part-of-fork:check-heights}
		\State \Return (\emph{false}, \emph{confirmStart})
		\EndIf
		\State \Return (\emph{true}, \emph{confirmStart})
		\EndFunction
	\end{algorithmic}
\end{algorithm}

\subsubsection*{Verifying Sufficient Block Confirmations}
Besides checking whether the header of the block supposedly containing the transaction to verify is part of the main chain of the source blockchain, it also needs to be ensured that the header is confirmed by enough succeeding block headers.


\cref{alg:is-confirmed} shows the pseudo code for this verification. Variable \emph{blockHash} indicates the block header from where to start the verification, and \emph{confirmations} specifies the required number of succeeding blocks. Note, the concrete number of block confirmations considered secure depends on the source blockchain, e.g., in Bitcoin six succeeding blocks are deemed sufficient~\cite{bitcoin-confirmations}.

The algorithm starts from the specified block header (i.e., \emph{blockHash}) and recursively calls the function for each succeeding block header each time decreasing the number of required confirmations by one~(Line~\ref{alg:is-confirmed:recursive-call}). In case the block header referenced by the parameter \emph{blockHash} does not exist, the algorithm returns false (Line~\ref{alg:is-confirmed:if-exists}f). Similarly, a block header that has not reached the end of the lock period yet may still be disputed and identified as invalid by disputers. Hence, in case the algorithm encounters a block header that is still locked, the verification fails as well~(Line~\ref{alg:is-confirmed:if-unlocked}f).

If the block header exists and is unlocked, it is checked whether further confirmations are required for the block header (Line~\ref{alg:is-confirmed:check-confirms}f). If no more confirmations are required, the function returns \emph{true}, since the block exists, is unlocked and confirmed. Otherwise, we check whether the header is a branch head, i.e., has no children~(Line~\ref{alg:is-confirmed:check-succs}f). If the header is a branch head but requires further confirmations, \emph{false} is returned, since the header has no succeeding blocks confirming it. If not, \Call{isConfirmed}{} is called on the next child. 

Notably, the algorithm always chooses the first child for the next recursive call~(Line~\ref{alg:is-confirmed:select-successor}). Of course, in case the block header is a branch junction, multiple children exist. Thus, the caller of the algorithm needs to make sure that the function is only called on a block header and a number of confirmations so that no further branch junction that is unlocked is reached as the algorithm would need to choose the right child in that case. 

To avoid this edge case, the function \Call{isPartOfMainChain}{} presented in \cref{alg:part-of-fork} precalculates the starting block header for the confirmation verification (variable \emph{confirmStart}), i.e., since it already traverses all branch junctions between the main chain's head and the requested block header, it can store any branch junction it traverses that is already unlocked. Since all predecessors of that branch junction are ensured to be unlocked as well, we can start the confirmation verification from that branch junction instead of from the requested block header with a reduced number of confirmations. If we encounter another branch junction during the confirmation verification, the branch junction will be locked, stopping the verification. If the branch junction were not locked, the branch junction would have already been returned by function \mbox{\Call{isPartOfMainChain}{}} as starting point for the confirmation verification. 

\begin{algorithm}[t]
	\caption{Verifies whether the block referenced by \emph{blockHash} has at least as many confirmations as specified in the parameter \emph{confirmations}.}
	\label{alg:is-confirmed}
	\begin{algorithmic}[1]
		\Function{isConfirmed}{\emph{blockHash, confirmations}}
		\If {\emph{headers}.contains(\emph{blockHash}) == \emph{false}}\label{alg:is-confirmed:if-exists}
			\State \Return \emph{false}
		\EndIf
		\State \emph{header} = \emph{headers}.get(\emph{blockHash})\label{alg:is-confirmed:get-header}
		\If {\emph{header.m.lockedUntil} $>=$ \emph{now}}\label{alg:is-confirmed:if-unlocked}
			\Return \emph{false}
		\EndIf
		\If {\emph{confirmations} $==$ 0}\label{alg:is-confirmed:check-confirms}
			\Return \emph{true}
		\EndIf
		\If {\emph{header.m.chldn.length} $==$ 0}\label{alg:is-confirmed:check-succs}
			\Return \emph{false}
		\EndIf
		\State \emph{child} = \emph{header.m.chldn}[0]\label{alg:is-confirmed:select-successor}
		\State \Return \Call{isConfirmed}{child, confirmations-1}\label{alg:is-confirmed:recursive-call}
		\EndFunction
	\end{algorithmic}
\end{algorithm}

\subsubsection*{Verifying Merkle Proof of Membership}
After verifying that block $b$ is part of the source blockchain's current main chain and that block $b$ is confirmed by at least $n$ succeeding blocks, the Testimonium relay checks the Merkle proof of membership. A Merkle proof depends on the data structures used in the source blockchain to represent Merkle trees. Our proof of concept~(see \cref{sec:evaluation}) implements Merkle proof of memberships for Merkle Patricia Tries~\cite{merklepatriciatrie}, a Merkle tree variation used in blockchains like Ethereum and Ethereum Classic.

If the verification of the Merkle proof fails, $tx$ is not part of block $b$. If it is successful, $tx$ is included within the block. Since the Testimonium relay has already verified block $b$'s membership in the main chain of the source blockchain and $n$ blocks succeeding $b$, it can be assumed that transaction $tx$ is in fact part of the source blockchain. 

As mentioned before, by specifying a sufficiently large number of confirmations, clients requesting a verification increase the probability that transaction $tx$ remains in the main chain of the source blockchain. Further, the verification procedure relies on the source blockchain's headers being exactly replicated within the Testimonium relay on the destination blockchain. Therefore, it is important that the disputers challenge any illegal block headers during the lock period. The details of disputing block headers are discussed in the next section.

\subsection{Disputing Block Headers}
\label{sec:disputing-headers}
To keep the cost of header submissions low, the Testimonium relay optimistically accepts newly submitted block headers without performing the source blockchain's header validation procedure at first. While this potentially enables illegal block headers to enter the relay, transaction inclusion verifications on such illegal block headers are prevented in Testimonium by means of a validation-on-demand pattern.

That is, each newly submitted header is assigned a lock period during which off-chain clients (i.e., the disputers) can dispute any block headers they deem illegal. When a block header is disputed, the header validation according to the validation protocol of the source blockchain is carried out. If the validation fails, the illegal branch, i.e., the block header together with all its descendants is eliminated from the Testimonium relay. Only after the lock period has passed, transaction inclusion verifications on the submitted headers can be carried out.

The validation-on-demand pattern can be leveraged whenever the block header validation is so costly that validating every single block header becomes too expensive.

\begin{algorithm}[t]
	\caption{Performs the validation of the disputed block header. If the validation fails, the header and all its descendants are removed. The function returns a list containing the addresses of submitters of illegal headers.}
	\label{alg:dispute}
	\begin{algorithmic}[1]
		\Function{disputeHeader}{\emph{blockHash}}
		\State \emph{header} = \emph{headers}.get(\emph{blockHash})
		\If {\emph{header.m.lockedUntil} $>$ \emph{now}} \label{alg:dispute:check-lock}
		\Return [] \label{alg:dispute:ret-empty-unlocked}
		\EndIf
		\If {\Call{isValid}{\emph{header}} == \emph{true}} \label{alg:dispute:ext-val}
			\Return [] \label{alg:dispute:ret-empty-valid}
		\EndIf
		\State \emph{submitters} = \Call{pruneBranch}{\emph{blockHash}} \label{alg:dispute:call-prunefork}
		\State \emph{parent} = \emph{headers}.get(\emph{header.parentHash})
		\State \emph{parent.m.chldn}.remove(\emph{blockHash}) \label{alg:dispute:remove-from-parent}
		\If {\emph{parent.m.chldn.length} == 0} \label{alg:dispute:check-head}
			\State \emph{branchHeads}.add(\emph{header.parentHash}) \label{alg:dispute:add-parent-to-epts}
		\EndIf
		\If {\emph{parent.m.chldn.length} == 1} \label{alg:dispute:check-1-succ}
			\State \Call{updateDesc}{parent.m.chldn[0], \mbox{parent.m.junction, parent.m.branchId}} \label{alg:dispute:set-junction-bid-succs}
		\EndIf
		\State \emph{mainChainHead} = \Call{getMainChainHead}{\,} \label{alg:dispute:set-valid-fork}
		\State \Return submitters
		\EndFunction
	\end{algorithmic}
\end{algorithm}

\subsubsection*{Validation-on-demand}
\cref{alg:dispute} shows the pseudo code for the steps performed whenever a block header is challenged by a disputer. The algorithm takes the hash of the disputed header as input parameter. First, it checks whether the disputed header is still locked, i.e., whether the lock period has not expired yet~(Line \ref{alg:dispute:check-lock}). Headers which are not locked anymore are deemed valid, so they cannot be disputed. 

If the lock period of the header is still active, the header validation is triggered (Line~\ref{alg:dispute:ext-val}). The concrete implementation of this validation depends on the block header validation procedure of the source blockchain. In our reference implementation, the verification of Ethereum block headers is carried out on block header disputes, since verifying Ethash---the consensus algorithm used by blockchains such as Ethereum and Ethereum Classic---for every submitted block header would otherwise lead to high operational cost~(see~\cref{sec:evaluation}).

If the validation returns \emph{true}, i.e., the block header is valid and was falsely disputed, the function aborts (Line~\ref{alg:dispute:ret-empty-valid}). In case the disputed header is in fact invalid (the validation returns \emph{false}), the illegal branch originating from the disputed header is removed from the Testimonium relay. For that, the function \Call{pruneBranch}{} is called~(Line~\ref{alg:dispute:call-prunefork}). 

After the deletion of the illegal branch, the header's hash is also removed from its parent's child list~(Line~\ref{alg:dispute:remove-from-parent}). Since the child list has been changed, we check in Line~\ref{alg:dispute:check-head} whether the parent of the disputed block header has now become a branch head. If so, the parent's hash is added to the set of branch heads~(Line~\ref{alg:dispute:add-parent-to-epts}). If the parent has exactly one child left (Line~\ref{alg:dispute:check-1-succ}), the parent is no longer a branch junction. Hence, we set the fields \emph{m.junction} and \emph{m.branchId} of all descendants up to and including the next branch junction or branch head to the corresponding fields of the parent (call of \Call{updateDesc}{} in Line~\ref{alg:dispute:set-junction-bid-succs}). We omit the pseudo code of this simple function due to space constraints. Further, the removal of an entire branch from the Testimonium relay may change the main chain. Thus, in Line~\ref{alg:dispute:set-valid-fork}, the head of the main chain is recalculated.

\subsubsection*{Pruning Illegal Branches}
\cref{alg:prune} shows the pseudo code for the \Call{pruneBranch}{} function which is called to remove an illegal branch. Before the disputed block header itself is removed, the function is called recursively for each child of the disputed block header~(Line~\ref{alg:prune:recursive-call}). The recursive invocation of \Call{pruneBranch}{} stops in case the header referenced by \emph{blockHash} has no children, i.e., it is a branch head. If so, besides removing the header from the \emph{headers} hash map, the header's hash is also removed from the set of branch heads (Lines~\ref{alg:prune:remove}-\ref{alg:prune:remove-endpt}).

 
To make submitters of illegal block headers accountable, the \Call{disputeHeader}{} function returns a list containing the addresses of the submitters of all the block headers being removed with the illegal branch. Of course, in case no header was successfully disputed (e.g., due to an expired lock period), no address is returned. The returned addresses are used to penalize submitters to discourage the submission of illegal block headers in the future (see \cref{sec:incentive}).

The correct functioning of the Testimonium relay is only ensured if submitters continuously submit block headers of the source blockchain to the destination blockchain and if disputers dispute submitted illegal block header. However, clients that submit and dispute block headers incur cost. Thus, an incentive structure for encouraging participation is needed. The details of this incentive structure are explained in the next section. 
 
\begin{algorithm}[t]
	\caption{Prunes the branch starting at the header referenced by \emph{blockHash}}
	\label{alg:prune}
	\begin{algorithmic}[1]
		\Function{pruneBranch}{\emph{blockHash}}
		\State \emph{header} = \emph{headers}.get(\emph{blockHash})
		\State \emph{submitters} = []
		\ForAll {\emph{child} in \emph{header.m.chldn}} \label{alg:prune:for-start}
			\State \emph{branchSubmitters} = \Call{pruneBranch}{\emph{child}} \label{alg:prune:recursive-call}
			\State \emph{submitters}.appendAll(\emph{branchSubmitters})
		\EndFor \label{alg:prune:for-end}
		\State \emph{headers}.remove(\emph{blockHash}) \label{alg:prune:remove}
		\State \emph{submitters}.append(\emph{header.m.submitter}) \label{alg:prune:append-submitter}
		\If {\emph{branchHeads}.contains(\emph{blockHash})}
			\State \emph{branchHeads}.remove(\emph{blockHash}) \label{alg:prune:remove-endpt}
		\EndIf
		\State \Return submitters
		\EndFunction
	\end{algorithmic}
\end{algorithm}

\subsection{Incentive Structure}
\label{sec:incentive}
 Without an incentive structure that compensates off-chain clients for submitting and disputing block headers, submitters and disputers may have no interest in participating. The incentive structure we propose rewards off-chain clients for submitting and disputing block headers and also discourages submitters from submitting illegal block headers.

To compensate the disputers for challenging illegal block headers, the submitters are required to deposit a stake. The stake is locked for the duration of the lock period of newly submitted block headers. While the stake is locked, it cannot be withdrawn and cannot be used for submitting further block headers. After a submitted header has passed the lock period without a dispute, the submitter gets back control of the corresponding locked stake. However, in case the block header is disputed successfully within the lock period, i.e., the validation of the block header fails, the disputer that triggered the dispute earns the locked stake of the submitter as well as any stake that was locked for any descendant of the illegal block header. Not only does this incentivize disputers, it also discourages submitters from submitting illegal block headers as they risk losing the deposited stake. Of course, disputers are only incentivized to dispute headers if the potential reward is higher than the cost of executing the dispute.

To encourage the submission of block headers, submitters receive a fee every time their submitted headers are used for verifying the inclusion of transactions. This verification fee is paid by the client requesting the verification. In order to fully compensate submitters, the total fees earned by transaction inclusion verifications on each header need to be greater than the initial submission cost for that header~(\cref{eq:fee-condition}).
 
\begin{equation}\label{eq:fee-condition}
	\textit{fee} \times \textit{no. of verifications} > \textit{submission cost}
\end{equation}

The minimum verification fee can thus be calculated as the submission cost of a block header divided by the number of verifications taking place on the submitted block header (\cref{eq:fee}).

\begin{equation}\label{eq:fee}
	\textit{fee} > \frac{\textit{submission cost}}{\textit{no. of verifications}}
\end{equation}

Applying a validation-on-demand pattern together with an incentive structure rewarding participation already lowers the cost of submitting new block headers to the relay. However, submission cost can be decreased even further by applying a slightly modified version of the content-addressable storage pattern~\cite{eberhardt2017or} which is explained in the next section.\\\\

\subsection{Further Optimization}
\label{sec:optimization}
So far, we have assumed that all data needed for disputes and transaction inclusion verifications is directly stored in the smart contract implementing the Testimonium relay. However, in blockchains like Bitcoin or Ethereum, all submitted transactions including their parameters are implicitly recorded in each blockchain's transaction history. We can take advantage of this fact storing only the hash of the block header, the block number and certain meta data in the smart contract itself. Fields such as the parent hash or the Merkle root hash no longer need to be kept in the smart contract, thus reducing the amount of stored data per block header which subsequently reduces submission cost. 

Whenever clients initiate a dispute or a transaction inclusion verification, they read the required full header data from the corresponding submit-transactions recorded in the transaction history and provide it to the smart contract. The contract can then verify the provided headers' integrity by recalculating their hashes and comparing it to the hashes stored in the smart contract. This way, no trust in the client invoking a transaction inclusion verification or a block header dispute is required.

While this pattern further reduces submission cost, it leads to an increment of cost for transaction inclusion verifications due to the increased amount of data that has to be passed along with each transaction. In the next section, we provide an extensive evaluation of Testimonium assessing this trade-off as well as providing a comprehensive security analysis. 

\section{Evaluation}
\label{sec:evaluation}
In this section, we evaluate the proposed relay scheme with regards to operational cost and security. Further, we look at the prerequisites that must be fulfilled to deploy a Testimonium relay scheme.

\subsection{Quantitative Analysis}
\label{sec:quantitative}
The advantages of Testimonium over traditional blockchain relays become apparent when the execution of the source blockchain's standard header validation on the destination blockchain is very costly. One example of this would be a bi-directional relay between Ethereum and Ethereum Classic. Both, Ethereum and Ethereum Classic, use Ethash as \ac{PoW} algorithm and the \ac{EVM} as execution environment. However, for the verification of Ethash, no native opcodes exist in the \ac{EVM}. Even with gas-optimized implementations as the one provided by SmartPool~\cite{luu2017smartpool}, the verification of Ethash for a single block header still costs around 3~million gas, potentially leading to very high operational cost when fully validating every single block header. Hence, these blockchains are a perfect fit for the validation-on-demand pattern employed by the Testimonium relay scheme. 

A further advantage of implementing Testimonium first for Ethereum-based blockchains is in the context of sidechains. Sidechains can be used to increase overall transaction throughput of a blockchain platform by outsourcing certain transactions to a sidechain~\cite{back2014enabling}. To transfer digital assets between the main chain and the sidechain, relays can be used to prove the existence of certain pieces of state on one chain from the other chain and vice versa. Ethereum being the most popular blockchain with regards to \acp{DApp} and digital assets~\cite{cai2018decentralized, diangelo2019survey} has experienced severe scalability issues in the past~\cite{cai2018decentralized}. Sidechains have been proposed to combat these issues~\cite{poon2017plasma}. As Testimonium leads to cost savings of up to 92\% over traditional relays for Ethereum-based blockchains, the operation of sidechains for Ethereum becomes a lot cheaper.

\subsubsection*{Evaluation Setup}
To evaluate the operating cost of Testimonium, we have implemented a total of three prototypes (\emph{Testimonium1}, \emph{Testimonium2}, and \emph{Baseline}, respectively) for \ac{EVM}-based blockchains like Ethereum and Ethereum Classic. The prototypes \emph{Testimonium1} and \emph{Testimonium2} both implement the validation-on-demand pattern. \emph{Testimonium2} additionally applies the content-addressable storage optimization as explained in \cref{sec:optimization}. The third prototype \emph{Baseline} acts as baseline for our experiments. This prototype does not implement the validation-on-demand pattern. Instead, it fully validates each block header at submission as done by traditional relays such as BTCRelay. Furthermore, \emph{Baseline} does not use the optimized search algorithm for transaction inclusion verifications and instead implements a na\"{i}ve search starting from the main chain's head, traversing each header until the header supposedly containing the transaction is found or the genesis block is reached. The functionality of each prototype is summed up in~\cref{tb:prototypes}.

\begin{table}[]
\centering
\caption{Prototype Functionality}
\label{tb:prototypes}
\begin{tabular}{@{}l*{2}{p{0.3cm}}l@{}}
\textbf{Functionality}     & \rotatebox{45}{\emph{Baseline}} & \rotatebox{45}{\emph{Testimonium1}} & \rotatebox{45}{\emph{Testimonium2}} \\ \midrule
Validation-on-submission   & \checkmark &            &              \\
Validation-on-demand       &            & \checkmark & \checkmark   \\
Content-adressable storage &            &            & \checkmark   \\
Na\"{i}ve search           & \checkmark &            &              \\
Optimized search            &            & \checkmark & \checkmark   \\ 
\end{tabular}
\end{table}

\begin{figure*}[t]
	\includegraphics[width=\linewidth]{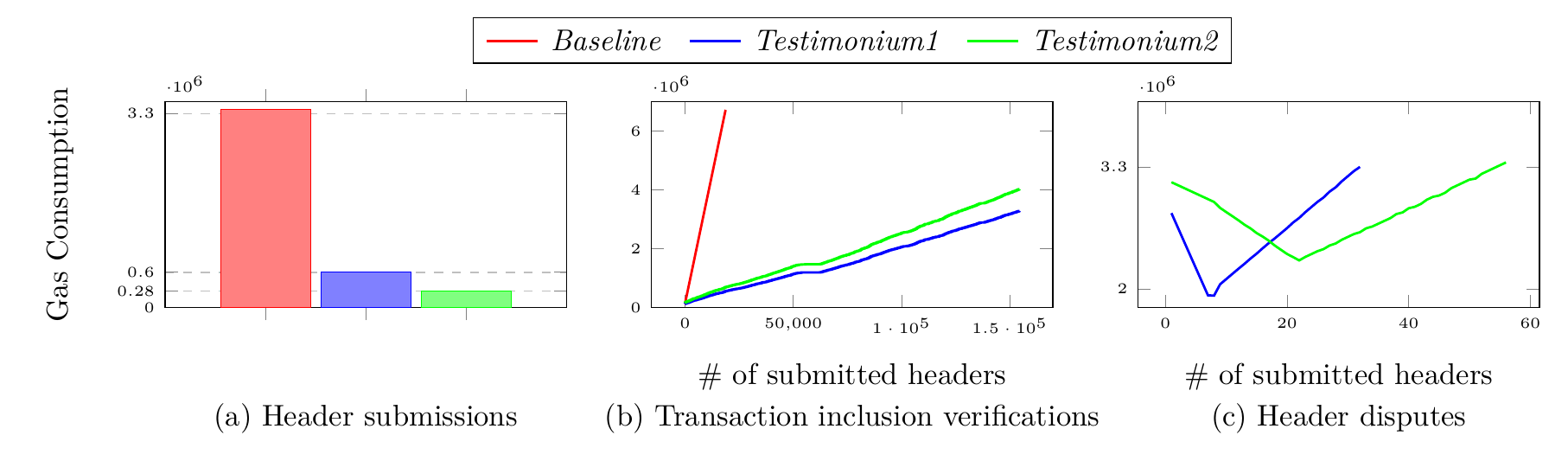}
	\caption{Gas consumption of the Testimonium relay}
	\label{fig:results}
\end{figure*}

Using a Geth light client (version 1.9.10), we collected 154,445 block headers containing 2,542 branches from the Ethereum main network over a period of two months. Note that we also count uncle blocks as branches, since---when submitted to Testimonium---they would introduce a new branch. We then fed these block headers into the three prototypes that were deployed as smart contracts to a private development blockchain running on a Parity Ethereum node~(version~2.6.8-beta,~--config dev). All three prototypes were initialized with block \#9121452 as genesis block.

A fully functional reference implementation of all concepts and algorithms of Testimonium, an off-chain client written in Go, and the evaluation are available as open-source projects on GitHub\footnote{\url{https://github.com/pantos-io/testimonium}}$^{,}$\footnote{\url{https://github.com/pantos-io/go-testimonium}}$^{,}$\footnote{\url{https://github.com/pantos-io/testimonium-evaluation}}. For repeatability, the evaluation project not only contains the three prototypes used for the evaluation, but also the evaluation scripts, the necessary block header data as SQL dump, and the results.

\subsubsection*{The Experiments}
In the first experiment, we analyze the operational cost (i.e., the cost of submitting block headers), the cost of verifying transaction inclusions, and whether the source blockchain is correctly replicated within Testimonium. For that, we continuously submit all block headers of our dataset to each prototype. The headers are submitted in ascending order according to their block numbers and timestamp fields. After each submission, a transaction inclusion verification on the genesis block (block \#9121452) is triggered. Since the replicated header chain within Testimonium grows after each submission, the algorithm checking whether block \#9121452 is part of the main chain has to deal with a growing number of headers. This allows us to observe the cost of transaction inclusion verification with an increasing search depth.

We determine the cost of each operation by measuring the gas consumption of its corresponding Ethereum transaction. Furthermore, after each submission, we log the head of the main chain and the currently submitted header's branch id and branch junction within Testimonium. This allows us to verify whether the submitted headers of the source blockchain (i.e., Ethereum main network) are correctly replicated within the prototypes running on the destination blockchain (i.e., private development blockchain). 


To measure the cost of header disputes, we repeat the first experiment, however, instead of performing a transaction inclusion verification after each submission, we trigger a dispute on the genesis block (block \#9121452). For simplicity, as we are primarily interested in the cost caused by the removal of branches, we remove the branch originating from the disputed header regardless of the actual result of the header validation. Additionally, after each dispute, all removed headers are resubmitted to restore the state as it was before the dispute. This allows us to observe the dispute cost with a growing number of headers that have to be pruned. The cost of each dispute (in gas) is logged only for prototypes \emph{Testimonium1} and \emph{Testimonium2}, since prototype \emph{Baseline} already performs the full header validation at time of submission.

\subsubsection*{Results}
\label{sec:results}

\cref{fig:results} shows the results from the experiments. \cref{fig:results}(a) shows the average gas consumption per header submission for each prototype. With 612,348 gas (standard deviation of 6,592 gas), \emph{Testimonium1} achieves a significant cost reduction of 82\% over \emph{Baseline} (average gas consumption of 3,369,653 gas, standard deviation of 5,101 gas). By applying the content-addressable storage pattern, \emph{Testimonium2} reduces the average gas consumption of \emph{Testimonium1} by 54\%, resulting in gas cost of 284,041 for every submitted header (standard deviation of 3,679 gas). Compared to \emph{Baseline}, \emph{Testimonium2} reduces submission cost by 92\%.

\cref{fig:results}(b) depicts the cost for transaction inclusion verifications on the genesis block (block \#9121452) for each prototype. The x-axis denotes the number of succeeding block headers that have already been submitted to the relay. Since transaction inclusion verifications are always performed on the genesis block, the search algorithms verifying the membership on the main chain have to cope with an increasing search depth. Prototype \emph{Baseline} using the na\"{i}ve search algorithm reaches the private blockchain's block gas limit of 6.7 million gas already after 18,766 submitted headers. \emph{Testimonium1} and \emph{Testimonium2} can cope with the growing search depth at much lower cost. Notably, \emph{Testimonium2} is slightly more expensive than \emph{Testimonium1} due to the implementation of the content-addressable storage pattern requiring the full block header to be provided at every transaction inclusion verification. Hence, applying this pattern is a trade-off between low submission cost and slightly higher verification cost. Notably, gas consumption is measured in a worst-case scenario where each block header is submitted to the relay even if it may not be part of the actual main chain of the source blockchain. In practice, the verification cost measured for \emph{Testimonium1} and \emph{Testimonium2} may be much lower since submitters may be reluctant to submit headers which are not part of the main chain since these headers will not yield a profit.

\Cref{fig:results}(c) shows the dispute cost measured for \emph{Testimonium1} and \emph{Testimonium2} (note that \emph{Baseline} has no dispute cost at all, since it does not implement the validation-on-demand pattern). Despite the fact that a growing number of headers is removed with each conducted dispute, the dispute cost of both prototypes temporarily declines. This is because in Ethereum freeing up contract storage yields a so-called gas refund which is given at the end of a successful transaction execution~\cite{wood2014ethereum}. However, from a certain point on (after nine headers for \emph{Testimonium1} and 23 headers for \emph{Testimonium2}), the dispute cost starts to rise which is caused by the design rationale of Ethereum that the gas refund is capped up to a maximum of the half of the total gas consumed by a transaction~\cite{wood2014ethereum}. Furthermore, as shown in the figure, \emph{Testimonium1} reaches the block gas limit much earlier than \emph{Testimonium2}. Notably, the gas refund is given only after the successful execution of a transaction, i.e., reaching the block gas limit makes the transaction fail without yielding any gas refund. This is why both graphs stop at around 3,3 million gas (last successful disputes) before reaching the block gas limit. If a branch is too long to be disputed within a single invocation, the dispute function can be called multiple times whereas each invocation prunes one part of the entire illegal branch. 

In the first experiment, the branch id and branch junction of each submitted header as recorded by each of the three prototypes was logged. This data allows us to verify whether all branches of the dataset have been correctly replicated within the relays. In particular, we extracted all unique junctions from the results as well as from our dataset. A comparison of both lists shows that all 2,542 branches were correctly recognized by the three prototypes. In the next section, we analyze the security aspects of the proposed relay scheme.

\subsection{Security Analysis}
\label{sec:security}
This section provides an informal security analysis of Testimonium. We consider attacks on the relay scheme itself, e.g., relay poisoning, as well as blockchain-specific attacks. Further, we consider consequences on the relay scheme in case changes to the involved blockchains occur. 

For the following discussion, we suppose the set of off-chain clients (i.e., submitters and disputers) to remain static during an attack. Furthermore, our analysis is based on the following assumptions: (a)~no off-chain client is guaranteed to follow the protocol rules, (b)~the actions of many clients are driven by self-interest, and (c)~some clients may categorically deviate from the protocol. Accordingly, we categorize off-chain clients into three groups according to the \acs{BAR} (\acl{BAR}) model~\cite{Aiyer2003BARModel}. This model has found application in security analysis for blockchain protocols and extensions before~(e.g.,~\cite{herlihy19cross, Judmayer2019PayToWinIA}). Under this model, byzantine clients may depart arbitrarily from the protocol for any reason, e.g., they may be faulty or may just follow strategies optimizing an unknown utility function. Altruistic clients always follow the protocol rules, regardless of whether deviations would lead to a higher profit. They exhibit no adversarial behavior. Finally, rational clients are self-interested, aiming at maximizing their profit according to a known utility function. These clients will depart from the protocol if they expect doing so to yield a higher profit than being honest.

\subsubsection*{Relay Poisoning}
\label{sec:relay-poisoning}
When verifying the inclusion of transactions, the Testimonium relay relies on block headers submitted by off-chain clients. Thus, an attacker may try to poison the Testimonium relay with wrong information regarding the source blockchain. For that, the attacker must trigger a chain re-organization within the relay according to the consensus rules of the source blockchain, i.e., the attacker must submit enough block headers such that these headers form the new main chain within the relay. This would allow the attacker to perform transaction inclusion verifications on wrong information. For instance, an application relying on Testimonium could be tricked into performing actions on the basis of transactions that have never happened on the source blockchain. Essentially, an attacker can choose between two approaches. Either the attacker sends invalid block headers to the Testimonium relay or the attacker submits headers which are themselves valid but belong to blocks containing invalid transactions (i.e., a header validation in case of a dispute would not detect any anomaly). We discuss these two attack models in the next subsections.

\paragraph*{Incentive Attacks on Disputes} Option one to achieve relay poisoning is for the attacker to submit illegal block headers while preventing other disputers from disputing these headers. The advantage of this approach is that the attacker does not have to follow the source blockchain's consensus rules for creating block headers, e.g., the attacker does not have to solve the \ac{PoW} for each header. This enables the attacker to create block headers at a much faster rate. However, disputes of these illegal headers would be successful since the block header validation would inevitably fail. Hence, in order to launch a successful attack, the attacker needs to convince all participating disputers not to dispute any illegal headers for the duration of the headers' lock periods, e.g., by launching incentive attacks~\cite{Judmayer2019PayToWinIA}.

Imagine all disputers to act rationally. Since rational clients seek to maximize profit, they may deviate from the protocol if doing so yields a higher income. Thus, for the attack to be successful, the attacker needs to offer disputers an alternative that is more profitable than a successful dispute of illegal block headers. The more disputers are participating in the relay, the more expensive the attack becomes since each disputer needs to be convinced to follow the attack. However, if an adversary is able to convince all disputers, the attack is successful.

A more realistic scenario involves---in addition to rational clients---also altruistic clients. Since altruistic clients always follow the protocol rules, they will not join the attack. Subsequently, the attack inevitably fails since illegal block headers will always be disputed by altruistic disputers. As long as at least one altruistic disputer is participating, relay poising via incentive attacks on disputes is not possible.

\paragraph*{Incentive Attacks on Submissions} Option two for achieving relay poising is for the attacker to submit block headers that are valid according to the source blockchain's header validation procedure but belong to blocks containing illegal transactions. Disputing these headers would not be successful, since transactions cannot be validated by the relay. Hence, the attacker could perform transaction inclusion verifications on illegal transactions. If other submitters continue to submit the correct block headers from the source blockchain, the only way this attack can be successful is if the attacker is able to create and submit valid block headers at a faster rate than the network of the source blockchain, e.g., by launching a 51\% attack~\cite{narayanan2016bitcoin}. 

Alternatively, the attacker could try to convince the other submitters to refrain from submitting block headers for the duration of the attack. This way, the attacker's block headers would be the only ones arriving at the Testimonium relay on the destination blockchain. If all submitters act rationally, the attacker may be successful in convincing them to join the attack, e.g., by offering a bribe. However, since all submitters need to be convinced, the cost of this attack grows proportionally with the number of participating submitters, analogue to what has been discussed above. Again, in a scenario involving both altruistic and rational clients, at least all altruistic submitters will follow the protocol rather than join the attack. If there is at least one submitter continuously submitting the block headers created by the network of the source blockchain, the attacker's branch will not become the main chain within the Testimonium relay. Again, as long as at least one altruistic submitter is participating, the attack is not possible.

\subsubsection*{Blockchain-specific Attacks}
\label{sec:blockchain-attacks}
While the above sections discuss attacks targeting the Testimonium relay directly, an adversary may also attack Testimonium by exploiting security flaws of the underlying blockchain. For instance, to prevent honest clients from disputing illegal headers or from submitting new ones in order to poison the relay, an attacker may try to prevent transactions from being included in the destination blockchain~\cite{Judmayer2019PayToWinIA,mccorry2018smart,secbit2018Fomo3D,Teutsch2017}, or to tamper with already included transactions~\cite{Bonneau2016BABitcoin,Judmayer2019PayToWinIA,Liao2017WhaleTx,mccorry2018smart,Teutsch2017}. If an attacker is able to prevent disputes of illegal headers for the duration of the header's lock period, a poisoning attack may be successful. The same principle can be applied when aiming at preventing clients from submitting new block headers allowing the attacker to form the new main chain within Testimonium.

While this section highlights some blockchain-specific attacks possibly affecting the security of Testimonium, it does not make a claim for completeness. Rather, the purpose of this section is to point out that---even with altruistic clients participating---an attacker may be successful in attacking Testimonium by targeting the underlying blockchain. However, it is not a vulnerability of Testimonium per se when an attacker exploits security flaws of the blockchain executing Testimonium since an application can only be as secure as the system on which it is running.

\subsubsection*{Changes to the Source Blockchain}
As introduced in \cref{sec:design}, Testimonium keeps track of block headers of some source blockchain. Hence, besides deliberate attacks as discussed in the prior sections, changes to the source blockchain may affect the reliability of the Testimonium relay.



When introducing changes to the block header validation procedure of the source blockchain, the header validation becomes either less or more restrictive. In the first case, headers adhering to the new validation rules would be rejected by the Testimonium relay when being received, or disputers would be able to successfully dispute block headers that are actually valid under the new rules. If the header validation becomes more restrictive, newly introduced validation rules are not enforced by the Testimonium relay, possibly leading to the acceptance of headers illegal under the new rules. Thus, any change to the header validation procedure of the source blockchain requires an update of the header validation procedure performed by the Testimonium relay.

On the other hand, if changes to the source blockchain do not affect the header validation procedure, the Testimonium relay does not need to be updated. In case the community of the source blockchain does not reach consensus on an upcoming hard fork, the source blockchain may be split up, resulting in multiple instances of the same blockchain. Technically, such instances are hard forks originating at the same blockchain. While the Testimonium relay is able to keep track of forks occurring on the source blockchain, only one fork is used for verifying the inclusion of transactions. Hence, there may be a competition of multiple instances of the source blockchain to form the main chain within the Testimonium relay. Which fork eventually overtakes the others may be unclear at the time the blockchain is split up. If multiple blockchain instances were to be supported, additional deployments of Testimonium would be required. 

\subsection{Testimonium on other Blockchains}
\label{sec:bc-requirements}
The prototypes described in \cref{sec:quantitative} demonstrate the feasibility of the proposed blockchain relay for \ac{EVM}-based blockchains like Ethereum and Ethereum Classic. In this section, we analyze the requirements that must be fulfilled to deploy Testimonium on other blockchains.

\subsubsection*{Requirements for the Destination Blockchain}
When deploying a Testimonium relay on some destination blockchain, two requirements must be met. First, the destination blockchain must provide a scripting language expressive enough to implement the concepts outlined in \cref{sec:design} as well as the validation of the source blockchain's consensus algorithm (e.g., Ethash). In particular, the Testimonium relay on the destination blockchain must be able to execute the validation procedure without any constraint violations such as the block gas limit in Ethereum. Otherwise, modifications of the proposed algorithms may be necessary, e.g., to spread the validation of the consensus algorithm across multiple transactions.

Notably, it is sufficient to validate block headers only according to the source blockchain's consensus algorithm instead of performing a full header validation. As long as an attacker is not able to produce block headers adhering to the consensus algorithm at a higher rate than the network maintaining the source blockchain, block headers which are valid according to the consensus algorithm but still contain some illegal fields (e.g., wrong block height) will eventually not form the main chain within Testimonium since the majority of the network will not build upon these headers. Thus, even if these headers cannot be disputed, no transaction verifications are possible on these headers as long as the number of required confirmation is sufficiently high.

As shown in our reference implementation, Ethereum and Ethereum Classic are fully suitable for hosting Testimonium relays. Prominent examples of blockchains currently not satisfying this requirement are Bitcoin and its forks such as Bitcoin Cash and Bitcoin SV since their scripting language only features a limited set of operations~\cite{bitcoin-script}.

The second requirement for a destination blockchain is the compliance with certain security guarantees. In \cref{sec:blockchain-attacks}, we highlighted some blockchain-specific attacks that may break the security of Testimonium. Hence, Testimonium relays should only be implemented for destination blockchains on which such attacks are unlikely to be successful. An example of a blockchain providing very limited security is Expanse, since the low network hash rate of only $130$~GH/s at the time of writing\footnote{5 December 2019, \url{http://stats.expanse.tech/}} (as compared to Ethereum's $175$~TH/s)\footnote{5 December 2019, \url{https://etherscan.io/chart/hashrate}} poses significant risk of 51\%~attacks.

\subsubsection*{Requirements for the Source Blockchain}
To enable the Testimonium relay on the destination blockchain to verify (a) the membership of a block $b$ in the main chain of the source blockchain, (b) a sufficiently high number of confirmations, and (c) the inclusion of a particular transaction in~$b$, the source blockchain is required to follow the data structures proposed by Satoshi Nakamoto~\cite{nakamoto2008bitcoin}. In particular, to allow the Testimonium relay the verification of (a) and (b), the source blockchain needs to be a linked list of block headers where each header contains at least a hash pointer to the previous block and the block height. Furthermore, to check (c), each block header must store the hash of the root node of the Merkle tree containing the transactions. While blockchains like Bitcoin or Ethereum satisfy this requirement~\cite{narayanan2016bitcoin,wood2014ethereum}, ledgers such as IOTA do not since transactions are stored in a \ac{DAG}, the so-called Tangle~\cite{iota-tangle}. Last but least, a blockchain should only be used as source blockchain if the success of attacks on that blockchain (e.g., 51\% attack) is unlikely.

\section{Related Work}
\label{sec:related}
Testimonium enables blockchain interoperability by providing the means to exchange (transaction) data across the boundaries of blockchains. According to Buterin~\cite{buterin2016interoperability} blockchain interoperability comes in three flavors: hash-locking schemes, where operations on two blockchains are synced by the same initial trigger; notary schemes, where a trusted party ensures that information is transferred between blockchains; and relay schemes, where a source blockchain's block headers are replicated on a destination blockchain enabling the destination blockchain to verify that certain data exists on the source blockchain.

Hash-locking is used whenever clients want to synchronize actions across multiple blockchains or between multiple participants on a single blockchain, e.g., in atomic swaps~\cite{herlihy2018atomic} or payment channels~\cite{poon2016bitcoin}, respectively. To this end, hash-locking schemes require the synchronized activity of at least two participants on both blockchains. However, hash-locking cannot be used to transfer arbitrary information from one blockchain to another. 

Notary schemes (e.g.,~\cite{wanchain,cosmos,polkadot}) allow transfers of information between blockchains by relying on trusted parties acting as intermediate authorities, i.e., notaries. Notaries are solely responsible for transferring information from the source blockchain to the destination blockchain. As long as the information is signed by the majority of notaries, the information is regarded as truthful on the destination blockchain, i.e., besides verifying the notaries' signatures, no further on-chain verifications are necessary on the destination blockchain. This makes notary schemes relatively lightweight, but---due to the high degree of centralization---also prone to manipulation.


Testimonium itself is a blockchain relay solution with the most prominent comparative projects being BTC Relay~\cite{btcrelay}, PeaceRelay~\cite{luu2017peacerelay}, and Waterloo~\cite{waterloo1,waterloo2}. BTC Relay~\cite{btcrelay} was the first and so far only relay solution to be operational. Starting in 2016, it allowed relaying block headers from the Bitcoin blockchain to the Ethereum blockchain. Due to the specifics of Bitcoin, fully validating every submitted header on the Ethereum blockchain is feasible, however this is impractical for blockchains with expensive header validation protocols such as Ethereum. 

PeaceRelay~\cite{luu2017peacerelay} can be considered the first attempt of a relay for Ethereum-based blockchains. In its current implementation, PeaceRelay requires authorized clients to submit block headers. However, without on-chain validation or branch handling it is not decentralized and rather a notary scheme than a relay.

Waterloo~\cite{waterloo1,waterloo2} attempts to bridge the Ethereum and EOS blockchains. This project consists of two separate relay solutions, one from Ethereum to EOS and a second one from EOS to Ethereum. Similar to BTC Relay, both relays perform full header validations of the respective chains. The Ethereum validation on EOS implements the optimized Ethash validation, similar to the full validation reference implementation used in this paper. EOS uses delegated Proof of Stake as consensus mechanism, which by design allows cheaper on-chain computations and thus makes Ethash in this case feasible. For the other direction, delegated Proof of Stake relies on a changing set of block producers, which on average happens every 8 hours~\cite{waterloo1}. Consequently, for relaying EOS to Ethereum, it suffices to validate only those block headers where the block producers change. Waterloo performs the block header validation for every submitted block header. While this might be practical for a relay between Ethereum and EOS, this approach becomes problematic when executing the source blockchain's header validation on the destination blockchain is very expensive, e.g., for a relay between Ethereum and Ethereum Classic.

To conclude, a number of blockchain relays have been conceptualized so far. However, their applicability largely depends on the underlying blockchains. If fully validating a submitted block header is an expensive operation, these relays may become impractical due to high operational cost. If applied to relays between EVM-based blockchains both, BTC Relay and Waterloo, would face this problem. The Testimonium relay as proposed in this paper is able to overcome this challenge by applying a cost-efficient validation-on-demand pattern together with a sophisticated incentive structure encouraging participation.

\section{Conclusion}
\label{sec:conclusion}
In this paper we introduced Testimonium, a secure and cost-efficient blockchain relay scheme. A Testimonium relay is especially suitable for blockchains where fully validating every submitted block header is an expensive task. Our evaluation using a proof of concept implementation of Testimonium for \ac{EVM}-based blockchains showed that Testimonium is able to reliably verify the inclusion of transactions across blockchains while reducing the operational cost in comparison to traditional blockchain relays by up to 92\% without jeopardizing decentralization.

In the current approach, every block header of the source blockchain needs to be submitted to the destination blockchain even though only a few block headers might actually be used for transaction inclusion verifications. In future work, in an effort to reduce submission cost even further, we will investigate possible batch submission of block headers as well as look into optimization opportunities for the block header search algorithm. Finally, Testimonium will act as basis for the development of an atomic-commit protocol for distributed transactions between multiple blockchains.\\\\



\balance


\bibliographystyle{abbrv}
\bibliography{ms}  


%
%
%
%
\todo{Werden wir dieses ``Additional Authors''-Kapitel irgendwie los?}

\end{document}